\documentclass[useAMS,usenatbib]{mn2e}

\usepackage{graphicx} 
\usepackage{epsfig}
\usepackage{amsmath} 
\usepackage{txfonts}

\def\refe@jnl#1{{#1}}
\def\aj{\refe@jnl{Astron.~J.}}                  
\def\araa{\refe@jnl{Annu.~Rev.~Astron.~Astrophys.}}
\def\apj{\refe@jnl{Astrophys.~J.}}                 
\def\apjl{\refe@jnl{Astrophys.~J.~Lett.}}          
\def\apjs{\refe@jnl{Astrophys.~J.~S.~S.}}          
\def\aap{\refe@jnl{Astron.~Astrophys.}}            
\def\mnras{\refe@jnl{Mon.~Not.~R.~Astron.~Soc.}}   
\def\prd{\refe@jnl{Phys.~Rev.~D}}        
\def\fcp{\refe@jnl{Fund.~Cos.~Phys.}}  
\def\physrep{\refe@jnl{Phys.~Rep.}}   
\def\physlett{\refe@jnl{Phys.~Lett.}}

\title[]{Sunyaev-Zel'dovich contribution in CMB analyses}
\author[N. Taburet et al.]{N. Taburet\thanks{E-mail:
    nicolas.taburet@ias.u-psud.fr},
    M. Douspis, N. Aghanim,
  \\ Institut d'Astrophysique Spatiale, Universit\'e Paris-Sud 11 \& CNRS (UMR 8617), B\^at.
  121, 91405 Orsay Cedex, France}
\begin{document}

\date{}

\pagerange{\pageref{firstpage}--\pageref{lastpage}} \pubyear{2009}

\maketitle
\label{firstpage}

\begin{abstract}
The Sunyaev-Zel'dovich (SZ) effect has long been identified as one of
the most important secondary effects of the Cosmic Microwave
Background (CMB). On the one hand, it is a potentially very powerful
cosmological probe providing us with additional constraints and on the
other hand it represents the major source of secondary fluctuations at
small angular scales ($\ell \geq 1000$). We investigate the effects of
the SZ modelling in the determination of the cosmological
parameters. We explore the consequences of the SZ power spectrum
computation by comparing three increasingly complex {modelling},
from a fixed template with an amplitude factor to a calculation
including the full cosmological parameter dependency. We also examine
the dependency of the cosmological parameter estimation on the
intra-cluster gas description used to calculate the SZ spectrum. We
show that methods assuming an SZ template bias the cosmological
parameters (by up to 2$\sigma$ on $\sigma_8$) when the cosmology used
in the template deviates from the {reference} one. A joint CMB-SZ analysis
with a full cosmological dependency of the SZ spectrum does not suffer {from}
such biases and moreover improves the {confidence intervals} of $\sigma_8$ and
$\Omega_{\rm dm}h^2$ (2.5 and 2 times respectively) with respect to {a}
pure CMB analysis. However, {the latter method} is quite sensitive to the intra-cluster gas
parameters and hence requires extra information on the clusters to
alleviate the induced biases.

\end{abstract}


\section{Introduction}

Over the last decade, the low multipole observations of the Cosmic
Microwave Background (CMB) angular power spectrum, in intensity and
polarisation, show that a $\Lambda$CDM concordance model describes
accurately the Universe \citep{KomatsuWMAP09} with the basic
cosmological parameters constrained \citep{DunkleyWMAP09} with a
precision of the order of a percent. To obtain even better constraints
on ''standard'' cosmological parameters and in order to { further
  constrain the cosmological model (dark energy, running of the
  spectral index, etc.)}, various experiments are {collecting} data
especially {at high multipoles} (CBI, BIMA, ACBAR, SZA)
\citep{Holzapfel2000,Dawson2001,Padin2001,Runyan2003,Muchovej2007}. These experiments discussed the existence of a
power excess at large $\ell$ values that could be accounted for by point
sources, SZ effect or more exotic physics -non standard inflation,
primordial voids, {features in the primordial spectrum, primordial
  non-Gaussianity, etc}-
\citep{Elgaroy01,Cooray02alternativsmallscaleexcess,Griffiths03,CBIexcess05,BIMAfinal,marian06,Acbar08,CBI09,SZarray09}.
This highlights the necessity to carry out a consistent analysis of
the CMB signal, based on a model that describes {both} primary anisotropies
{and} also the secondaries arising from the interaction of CMB photons
with matter between the last scattering surface and the observer.

In the case of experiments that have access to high multipoles
($\ell \geq 1000$), the secondary SZ anisotropies and the point
sources contribution will dominate over the primary CMB. {A joint
  analysis of the CMB and SZ power spectra is thus
  necessary. Different parameterisations of the SZ spectrum have
  already been used to analyse the data from WMAP, CBI, ACBAR, and
  BIMA
  \citep{Spergel03,CBIexcess05,marian06,Spergel07,Kuo07,Acbar08,DunkleyWMAP09,CBI09}. In
  the present analysis, we address the issue of the calculation of the
  SZ spectrum used to fit the data.  We examine the effect each SZ
  description (power spectrum and intra-cluster gas model)} induces on
cosmological parameter estimation in terms of accuracy and possible
biases.

In section 2, we {briefly introduce the thermal SZ effect and the
  calculation of its {power} spectrum. We then present in section 3 the
  different methods used to jointly fit the CMB + SZ data. We compare
  and discuss in section 4 {the accuracy of the parameter estimation} for each of
  these methods and the possible induced biases.}  We summarize our
results in section 5.  Throughout the study we assume a flat
$\Lambda$CDM cosmological model and use the WMAP5 cosmological
parameters \citep{KomatsuWMAP09} : {$\sigma_8=0.817$, $n_{\rm s}=0.96$,
$\Omega_{\rm m}=0.279$, $\Omega_{\rm b}=0.046$}, $h=0.701$.

\section{The Thermal SZ angular power spectrum}
\label{sec:method}

The Sunyaev-Zel'dovich (SZ) \citep{SZ72} is the main secondary
anisotropy source at arcminute scales. It is made of two terms : the
first is the thermal SZ (TSZ) due to the inverse Compton scattering of
the CMB photons off the hot electrons in the intracluster gas, and the
second is the kinetic SZ (KSZ), a Doppler shift due to the proper
motion of clusters with respect to the CMB. {The KSZ power
  spectrum is approximately 2 orders of magnitude smaller than the
  TSZ. We thus neglect its contribution in the present analysis.}  The
TSZ has a characteristic spectral signature, namely a decrease of the
CMB intensity in the Rayleigh Jeans part of the spectrum and an
increase in the Wien part that is due to energy transfer from the hot
intracluster electrons to the CMB photons. This characteristic
frequency signature is given, in the non-relativistic approximation,
by $f(x)=\left[x\frac{e^{x}+1}{e^x-1}-4\right]$, where
$x=\frac{h\nu}{k_{\rm B}T_{\rm e}}$.

{In the context of the analysis of CMB data, it is necessary to
take into account the contribution of the TSZ effect. The most
useful tool to account for this contribution as a function
of the angular scale on the sky is the TSZ power spectrum.
It }can be obtained either from hydrodynamical
simulations or {from} analytical calculations. {Each of these
  approaches has its own limitations and drawbacks.}  In the first
case, the amplitude of the SZ spectrum {as well as its shape are
  sensitive to the simulation characteristics. The box-size and
  resolution, or smoothing length, of the simulation affect the
  relative amplitudes on large and small scales respectively.} When
the {box-size} is too small, the number of massive clusters { is}
underestimated {and so is} the SZ power at large scales. { The resolution
  of the simulation, or the smoothing length,} artificially
{decreases} the SZ power at small angular scales
\citep[see for example][]{WhiteHernquistSpringel02}. The {physical model used to
  describe} the gas is an other {important source of alteration} of the
SZ spectrum. {For example,
it has been showed that preheating (due to energy
feedback from supernovae for instance) as well as radiative cooling, that depends on the gas metalicity \citep{DolagHansen05}, respectively
increases or decreases the amplitude of the TSZ spectrum by a factor of
2 \citep{daSilva01}. The influence of radiative cooling on the SZ spectrum has also been studied with an analytical treatment by \citet{ZhangWu03} and has
been shown to significantly affect the amplitude of the SZ spectrum.}

{In the second case,} the analytical calculation of the TSZ power
spectrum {is based on two major ingredients:} the {halo} mass
function, along with a {model for the intra-cluster gas distribution}
within these halos. {Neglecting the correlation between haloes {
    which is much smaller than the Poisson term or the CMB signal}
  \citep{KomatsuKitayama99},} the TSZ angular power spectrum can be
calculated {as in} \citet{KomatsuSeljak2002}:
\begin{equation}
\label{eq:SZ_powspec}
C_\ell^{\rm SZ}=f^2(x)\int_0^{z_\rmn{max}}dz\frac{dV_\rmn{c}}{dzd\Omega}\int_{M_{\rm
    min}}^{M_{\rm
    max}}dM\frac{dn(M,z)}{dM}\left|\tilde{y_\ell}(M,z)\right|^2.
\end{equation}
{The mass function $n(M,z)$ is given by the
  theoretical expression of Press \& Schechter (1972) or by fitting
  formulae to N-body numerical simulations \citep{ShethTormen99, Jenkins01, Warren06}.
{The different mass function can lead to different predictions}
in terms of the predicted numbers of halo, and thus may induce
differences in the power spectrum \citep{KomatsuSeljak2002}. } The shape of the TSZ
angular power spectrum depends on the intra-cluster gas distribution
and properties through the two-dimensional Fourier transform on the
sphere of the 3D radial profile of the Compton $y$-parameter for
individual clusters,
\begin{equation}
\tilde{y}_\ell=\frac{4\pi}{D_{\rm A}^2}\int_0^\infty
y_{\rm 3D}(r)\frac{\sin\left(\ell r/D_{\rm A}\right)}{\ell r/D_{\rm A}}r^2dr,
\end{equation}
where $y_{\rm 3D}(r)= \sigma_{\rm T}\frac{k_{\rm B}T_{\rm
    e}(r)}{m_{\rm e}c^2}n_{\rm e}(r)$. {Several models for the
  electronic distribution $n_{\rm e}(r)$ can be used. The most
  commonly used are the $\beta$-profile \citep{Cavaliere76} and the
  polytropic gas distribution \citep{KomatsuSeljak2001} and extensions
  of these two. {The mean gas mass fraction, $f_{\rm
      g}$ is taken 0.086/h \citep{Laroque06}.} As for the
  intra-cluster gas temperature $T_{\rm e}$, it is usually assumed to
  be isothermal and equal to the virial temperature : {$k_{\rm
      b}T_{\rm e}=\frac{G \mu m_{\rm p}M}{3r_{\rm vir}}$, where $\mu$
   is the mean molecular weight and $r_{\rm vir}$ is calculated using the formula in \citet{BryanNorman98}}.} 
The shape of the TSZ angular power
spectrum also depends on the {cosmological model} through the comoving
volume $V_{\rm c}$, the angular diameter distance $D_{\rm A}$ and the
mass function $n(M,z)$.

\section{SZ in CMB analysis}
\label{sec:SZinCMBanalysis}

{The estimation of cosmological parameters from the CMB would ideally require
a pure primary signal. However the measured CMB data will contain additional contributions,
of which the TSZ dominates.}
The TSZ characteristic frequency signature allows us in principle to
remove the secondary contribution of the detected galaxy clusters from
the CMB in multi-frequency experiments.  However, on the one hand the
residual SZ signal, if not taken into account in the analysis, biases
the cosmological parameters see \citep[see][]{Taburet09}, and on the
other hand taking it into account requires a good knowledge of the
selection function.  An alternative method {to extracting the SZ
  contribution and modelling the residuals} is to model the total CMB
plus TSZ spectra and to determine the cosmological parameters using
both primary and secondary signals. This is the {approach} used in
{all} the present high $\ell$ study.

{We can naturally fit the total signal $C_\ell^{\rm tot}$ with the
  expression
\begin{equation}
C_\ell^{\rm tot}=C_\ell^{\rm CMB}(\hat\theta)+C_{\ell}^{\rm SZ}(\hat\theta)
\label{eq:method3}
\end{equation}
with $C_{\ell}^{\rm SZ}(\hat\theta)$ given by Eq. (\ref{eq:SZ_powspec})
as in \citet{marian06}, {where $\hat\theta$ stands for the set of cosmological parameters
we want to determine}. The main advantage of this, hereafter method
3, is that it includes the full cosmological parameter dependency of
the SZ power spectrum. Parameterisations of the SZ power spectrum are
also used to fit the CMB data. They are very efficient from the point
of view of computation time since they do not require the full
calculation of the SZ spectrum at each step of the Monte Carlo Markov
Chains (MCMC). As a result, parameterisations of the SZ spectrum can reduce
the chains convergence time by a factor of the order of 2. However, the drawback is that they do not reflect the full
cosmological dependency of the SZ spectrum. Two methods fall in this
category. In the first one,} the total CMB power spectrum can be
fitted with
\begin{equation}
C_\ell^{\rm tot}=C_\ell^{\rm CMB}(\hat\theta)+A_{\rm
  SZ}C_{\ell}(\hat\theta_0),
\label{equ:method1}
\end{equation}
 hereafter method 1, where $A_{\rm SZ}$ is an amplitude factor
 multiplying {an SZ spectrum template $C_{\ell}(\hat\theta_0)$
   calculated analytically for a given cosmology {described by the
     set of cosmological parameters $\hat\theta_0$} and intra-cluster
   gas distribution, or obtained from a given numerical simulation
   (i.e. for a given cosmology and gas physics). {In this analysis
     the templates we used were those used by the WMAP team.}
   
The {last} method accounts for the main variations in the SZ spectrum amplitude,
with the cosmological
   parameters $\sigma_8$ and $\Omega_{\rm b}h$ following
   \citet{KomatsuSeljak2002}. In this method 2, the total CMB spectrum
   is fitted with
\begin{equation}
C_\ell^{\rm tot}=C_\ell^{\rm
  CMB}(\hat\theta)+\sigma_8^7\left(\Omega_{\rm b}h\right)^2C'_{\ell}(\hat\theta_0),
\label{equ:method2}
\end{equation}
where $C'_{\ell}(\hat\theta_0)$ is the SZ spectrum for a given cosmology and
intra-cluster gas distribution.}

{In the following (Sect. \ref{sec:szp}), we investigate the
  effects on the cosmological parameter estimation of the three
{methods introduced above}. For methods 1 and 2, we
  used the same SZ spectrum template as that used by the WMAP team
  \citep{Spergel07} and which is based on
  \citet{KomatsuSeljak2001,KomatsuSeljak2002}. For method 3, we
  computed the SZ spectra using the \citet{ShethTormen99} mass
  function and the spherical $\beta$-profile with $\beta=2/3$ for simplicity. Furthermore in
  Sect. \ref{sec:phys}, we study the effects of varying this profile on the cosmological
  parameter estimation.
  
\par\bigskip 

In order to compare the three different methods, we create {mock}
data, at 100 GHz, in the form of temperature and polarisation power
spectra containing primary CMB anisotropies and SZ contribution {
  from all clusters} using equation \ref{eq:SZ_powspec}. We did not
consider the polarisation induced by clusters since it is negligible
compared to the primary one \citep[e.g.][]{Liu05}.
{In this study, we do not analyse the CMB data produced from
  multi-frequency observations after component separation. This would
  require to monitor precisely the residual signal after component
  separation which is shown to mix all foregrounds and may differ from
  one component separation method to another. Such an approach was
  followed by J.-A. Rubino-Martin and gives similar results as ours
  (private communication). We have rather chosen to use the best
  channel for CMB study : the 100 GHz channel. In this channel,
  galactic foregrounds (free-free, dust emission, synchrotron
  radiation) will contaminate the CMB minimally as well as
  extragalactic radio and IR point sources.} We also consider a
\emph{Planck}-like {gaussian and uncorrelated} noise power spectrum
with a 9.5 arcminutes beam.}  We ran MCMC analyses, using the CosmoMC
code \citep{LewisBridle02} with a modified version of the CAMB code
\citep{CAMB2000} including a module that calculates the SZ power
spectrum with its full cosmological dependency. {We followed for this
  module the computations detailed in Sect. 2.}  Given the instrumental
beam {we considered}, we limited the MCMC analysis to multipoles
smaller than $\ell_{\rm max}=3500$.  We used a gaussian likelihood
function since the probability distribution function of the SZ
spectrum is well approximated by a gaussian for a \emph{Planck}-like
survey \citep{ZhangSheth07}.  The MCMC were carried out on the set of
parameters $\hat\theta$: $\Omega_{\rm b}h^2$, $\Omega_{\rm dm}h^2$,
the ratio of the sound horizon to the angular diameter distance $100
\times \theta_h$, the optical depth at reionisation $\tau$, the
spectral index $n_{\rm s}$, the {CMB normalisation} $A_{\rm s}$, and
the SZ normalisation factor $A_{\rm SZ}$ when method 1 is used. The
deduced parameters are $\Omega_\Lambda$, $\Omega_m$, $\sigma_8$,
z$_{\rm re}$, H$_0$ and the age of the universe. To ensure that the
MCMC runs have converged we used the convergence criterion introduced
by \citet{Dunkley05}.

\section{Results}

\subsection{Effects of the SZ modelling on cosmological parameter
  estimation.}\label{sec:szp}

{As a first step we examine the choice of the SZ template (used in
  methods 1 and 2). We thus} present the MCMC results, when the SZ
template deviates from the real SZ angular power spectrum (i.e. 
that used to produce the {mock} data).  To account for this possibility,
the SZ template we use to fit our simulated data is based on the one
used by the WMAP team \citep{Spergel07}. We normalized this template
so that its amplitude at $\ell=2000$ equals the amplitude of the SZ
spectrum employed to create the {mock} data. At such multipoles, the gas
properties affect more the SZ amplitude than does the cosmology. As a
result, this normalisation ensures that the main difference between
the SZ spectrum used to create the data, and the template used to fit
them in methods 1 and 2, is due to the cosmology.

\begin{figure}
\includegraphics[width=8cm]{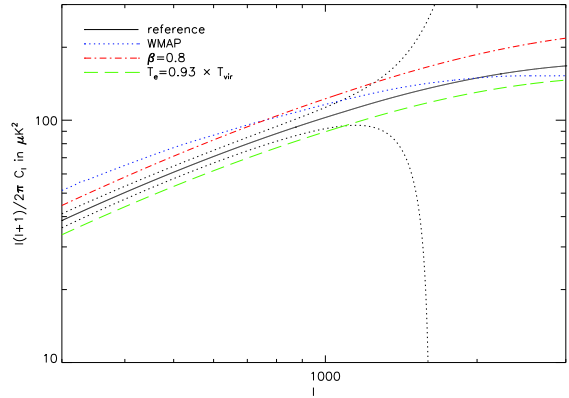}
  \caption{Theoretical SZ reference spectrum ($\beta=2/3$, $T_e=T_{vir}$, solid black
    line), SZ WMAP template (dotted blue line) normalised to the
    $\ell=2000$ value of the reference spectrum, SZ template of the
    reference model but with a {steeper} $\beta$ distribution of the
    intra-cluster gas ($\beta=0.8$, red dot-dashed line) and SZ
    spectrum when the electronic temperature is underestimated by
    7\%. The black dotted curves represent the $1 \sigma$ envelopes
    assuming a \emph{Planck} like experiment at 100 GHz with a 9.5 arcmin beam.}
\label{fig:SZtemplates}
\end{figure}

\begin{figure*}
\includegraphics[width=16cm]{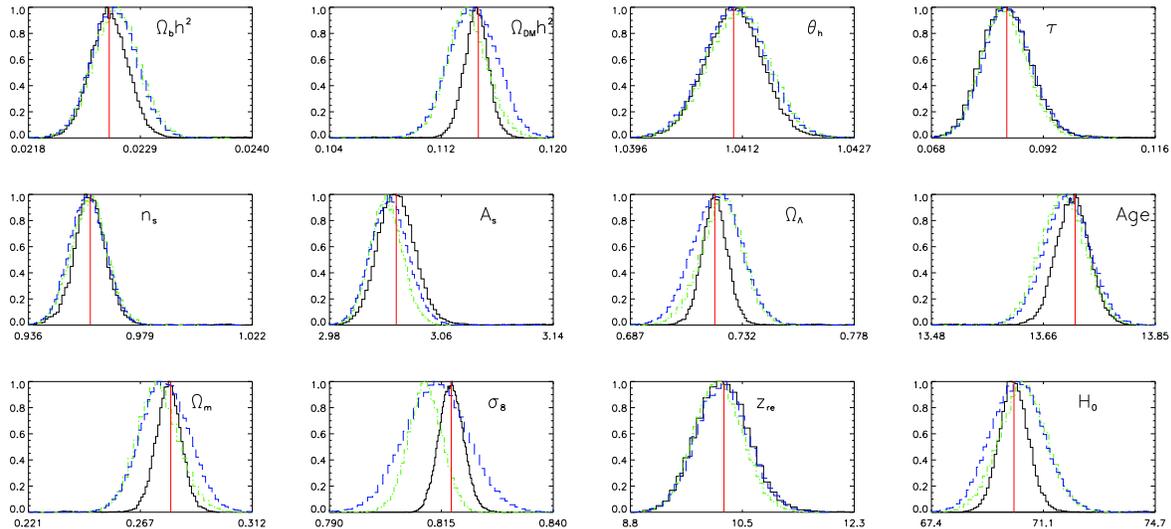} 
  \caption{One dimensional parameter distribution. The
    curves represent the distributions for the CMB+SZ signal fitted
    with the 3 different methods. Long dashed blue line : method 1, dot-dashed green line :
    method 2, solid black line : method 3. The SZ template
    used in method 1 and 2 (blue dotted line in figure
    \ref{fig:SZtemplates}) differs from that used to create the
    data (black line in figure \ref{fig:SZtemplates}). The vertical red lines represent the
    input values of the parameters we used to create our {mock} data. Note the substantial
    biases that can affect the estimation of $\sigma_8$ when using method 1 or 2.}
\label{fig:diffmethodecart}
\end{figure*}

We represent in figure \ref{fig:diffmethodecart} the one-dimensional
distribution of cosmological parameters {obtained with the three
  methods. Using the SZ parameterisation of method 1 (long dashed blue lines)
  enlarges the $1\sigma$ error bars on the parameters and biases the
  determination of some parameters. The values of $\Omega_{\rm dm}h^2$ and $A_{\rm
    s}$ are biased by a few tenths in units of the standard
  deviation. This translates, for the derived parameters, into a one-sigma
  bias on $\sigma_8$, and a few tenths in units of the standard
  deviation for $\Omega_\Lambda$, $\Omega_{\rm m}$ and $H_0$.}  These
biases {are due to the fact that} method 1{, by construction,}
allows to vary only the amplitude of the SZ power spectrum {and not its
  shape}.

When employing {the SZ parameterisation of method 2, the {confidence intervals are
improved}.
This can be explained by the fact that
  in method 1 we constrained an extra parameter ($A_{\rm SZ}$) while
  in method 2 the SZ amplitude is given by a combination of $\sigma_8$
  and $\Omega_{\rm b}h$.} However, as shown in figure
\ref{fig:diffmethodecart}, the biases on some cosmological parameters
    {become} more important. The values of $\Omega_{\rm b}h^2$, $\Omega_{\rm
      dm}h^2$ and $A_{\rm s}$ are biased by {a few tenths}
 in units of the expected precision {which, for the derived parameters, leads to a two-sigma
      bias on} $\sigma_8$, whereas the values of $\Omega_\Lambda$, $\Omega_{\rm m}$
    and $H_0$ {suffer from a one-sigma bias. These results can
      be explained by two facts~: first the shape of the SZ template in
      method 2 does not vary, and second its amplitude at
$\ell<2000$
      is higher than the SZ spectrum used to produce the data
      (Fig. \ref{fig:SZtemplates}). The latter forces the estimated
      $\sigma_8$ to be biased towards a low value in order to decrease
      the SZ amplitude (given by $\sigma_8^7(\Omega_{\rm b}h)^2$, see
      Eq. \ref{equ:method2} ) and thus to fit the data. For the same
      reason the $\Omega_{\rm b}$ and $H_0$ values are also biased. As for
      the second point, the ``fixed'' shape of the SZ template,
      deviating from the reference one, forces variations of $\Omega_{\rm
        m}$, $\Omega_\Lambda$ and $\Omega_{\rm b}$ so that the primary
      CMB spectrum compensates the shape modifications. }

\par
\begin{figure*}
\includegraphics[width=16cm]{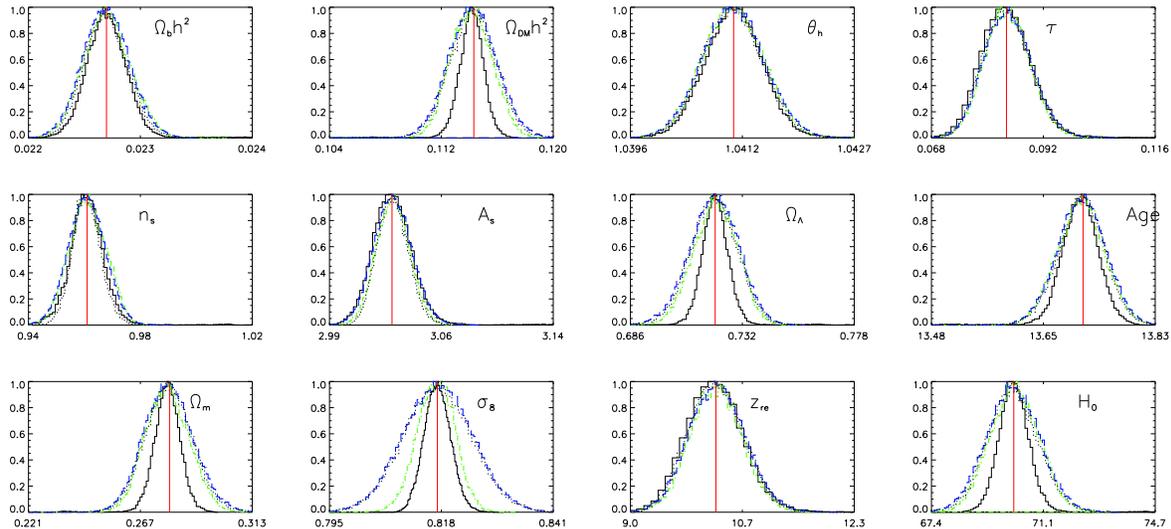}
  \caption{One dimensional parameter distribution. The coloured
    curves represent the distribution for the CMB+SZ signal fitted
    with the 3 different methods. Long dashed blue line : method 1, dot-dashed green line :
    {method 2, solid black line : method 3}. The SZ template
    used in method 1 and 2 is the same as that used to create the data. The black dotted
    line represents the parameter distributions when then signal
    contains only a pure primary CMB. The vertical red lines represent the
    input values of the parameters we used to create our {mock} data.}
\label{fig:diffmethodexact}
\end{figure*}

\bigskip
Method 3, which does not
require any SZ template with a fixed shape, gives unbiased
determination of the cosmological parameters as shown by the black curves in Fig.\ref{fig:diffmethodecart}.
The accuracy obtained with method 3 is also better than that obtained with methods 1 and 2. In order to
discuss this accuracy improvement, we now compare {method 3 to
  methods 1 and 2 using the SZ
template used to create the {mock} data.}  The one dimensional parameter distributions obtained
from the different methods are presented in figure
\ref{fig:diffmethodexact}. The black dotted lines represent the
parameter distribution when we fit with pure primary CMB a signal that contains primary CMB alone. As
expected, all 3 methods {give unbiased} parameters but their
associated error bars (i.e. the accuracy of the parameter
determination) differ.

Unsurprisingly, {adding an extra parameter (the SZ amplitude
  $A_{\rm SZ}$) in the MCMC analysis with method 1 enlarges all} the
error bars (Fig. \ref{fig:diffmethodexact} long dashed blue curves) that would be
obtained in a pure primary CMB analysis (black dotted line). {Method 2 (dot-dashed green
  curves) with its explicit and strong dependency of the SZ amplitude
  on $\sigma_8$ improves the constraints on $\sigma_8$, in comparison
  to the pure CMB analysis. At the same time,} the accuracy on the
spectral index $n_{\rm s}$ is deteriorated. {Increasing $n_{\rm
    s}$ increases the CMB power at high $\ell$ values
    but this can be
slightly compensated by a decrease of the SZ amplitude through
$\sigma_8$ or $\Omega_{\rm b}h$. The solid black curves represent the
  expected precisions obtained with method 3. We note that
  they are improved for all parameters except $n_{\rm s}$ as
  compared to the pure primary CMB analysis. {The improvement is}
particularly important for $\sigma_8$: {a factor 2.5 better}. The
accuracy on $\Omega_{\rm dm}h^2$ and on the derived parameters
$\Omega_\Lambda$, $\Omega_{\rm m}$, $H_0$ is also improved by a factor 2.

Method 3 provides better constraints on the cosmological parameters because
it allows to break degeneracies between models, as illustrated in figure \ref{fig:Cl_SZbreakdegeneracy}a.
This figure represents the {$68.3\%$ confidence region on}
parameters $\Omega_{\rm dm}h^2$ and $A_{\rm s}$ when
the analysis is only performed on these 2 parameters. An analysis
using only the primary CMB signal gives the distribution represented
by the red/dark ellipse. Using our SZ module to carry out an analysis {on the
SZ alone power spectrum} (black ellipse) shows that $A_{\rm s}$ and
$\Omega_{\rm dm}h^2$ are strongly degenerated but very differently with
respect to a pure CMB, which means that the SZ contains some
information complementary to that included in a pure CMB
signal. The analysis of the CMB + SZ signals based on method 3 results in the two
dimensional distribution represented by the green/light ellipse that is
smaller than the red one, taking advantage of the
information included in the SZ power spectrum.

\begin{figure*}
\begin{minipage}{.49\linewidth}
\centering
\includegraphics[width=8cm]{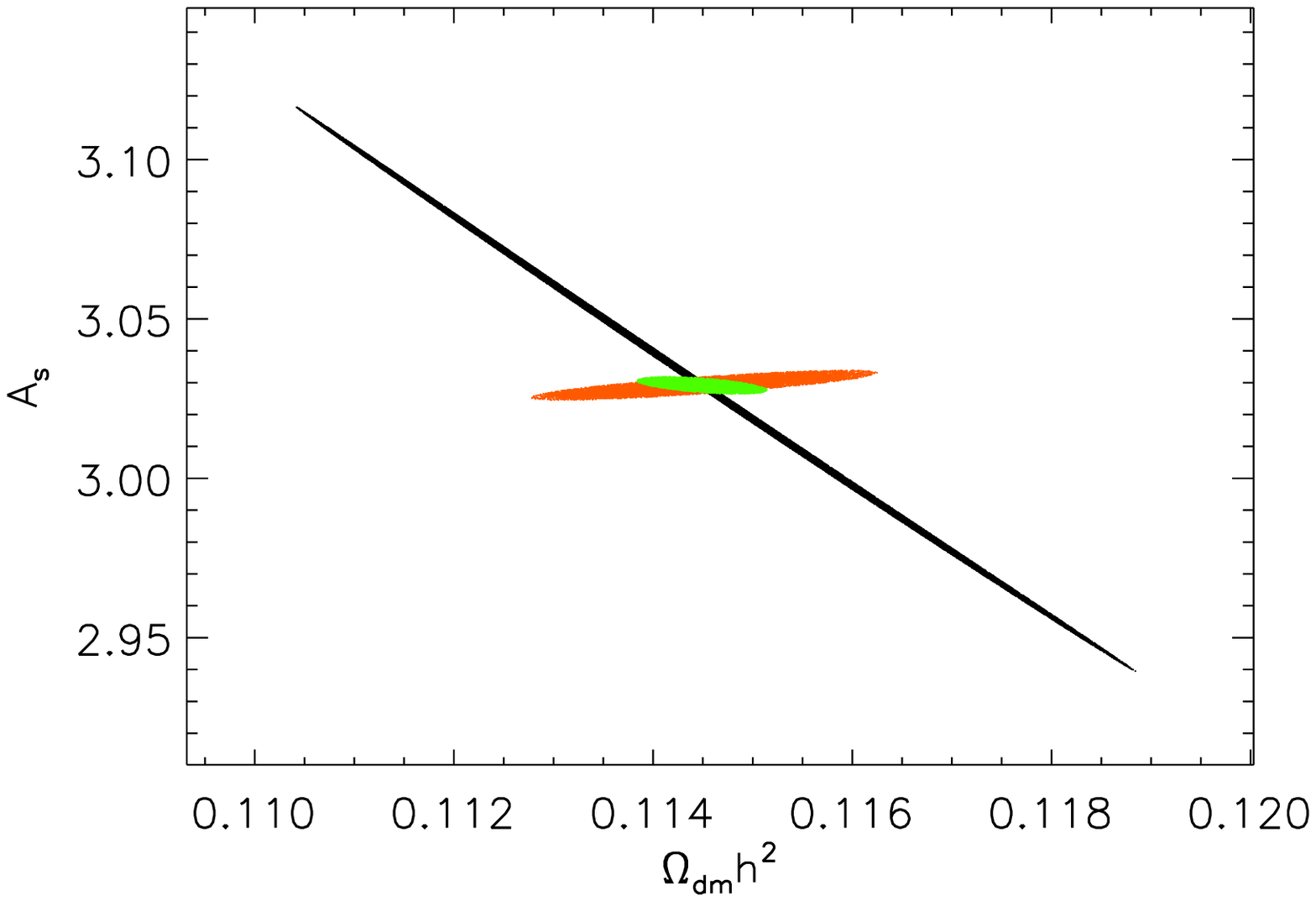}
\end{minipage}
\begin{minipage}{.49\linewidth}
\centering
\includegraphics[width=8cm]{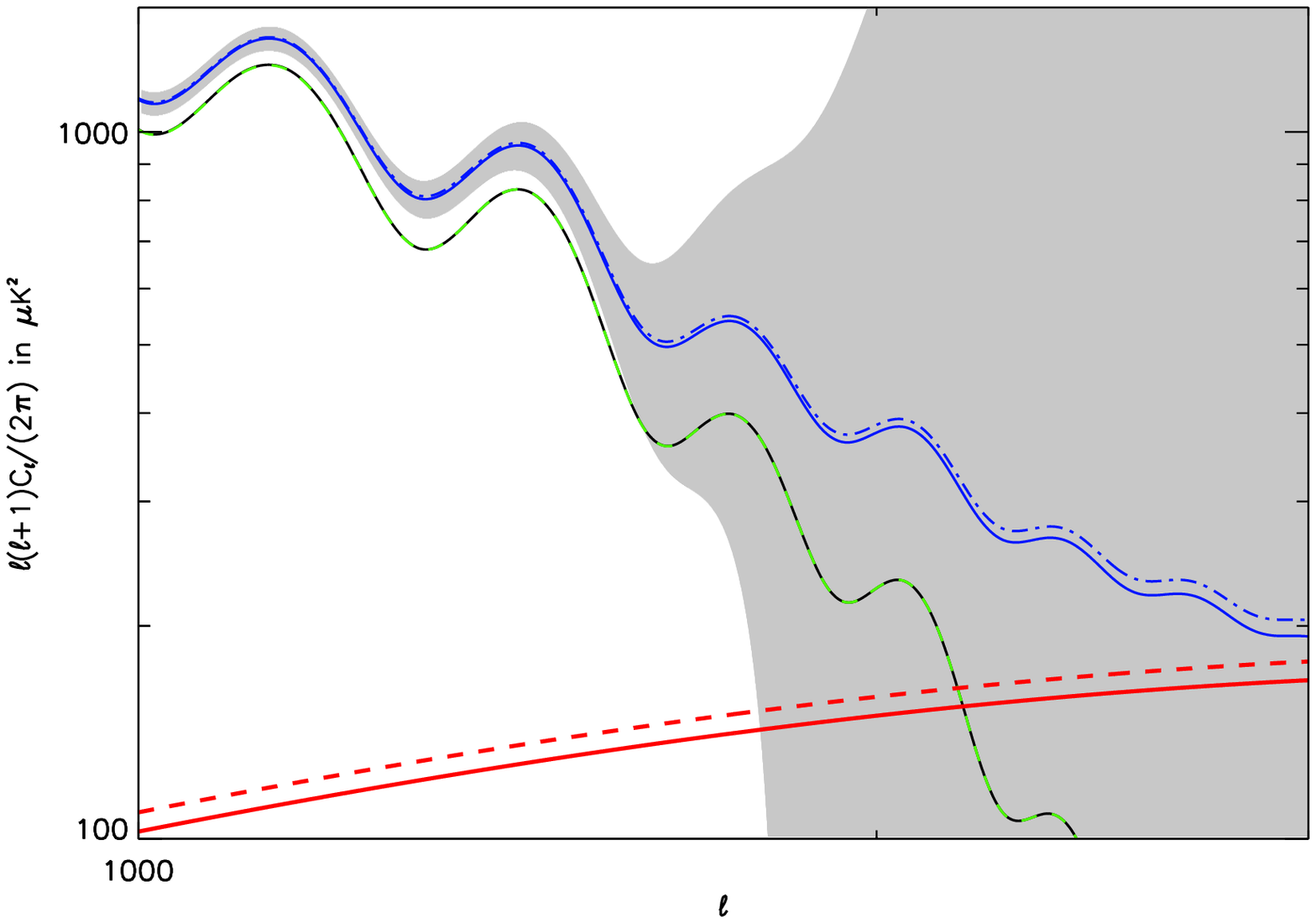}
\end{minipage}
  \caption{Left panel: Two dimensional parameter distribution. The black ellipse
    represents the $\Omega_{\rm DM}h^2$ / $A_{\rm s}$ {pairs} obtained
    from the Markov chains analysis of the pure SZ spectrum. The orange (dark grey)
    ellipse represents the {pairs} obtained from the analysis of the
    pure CMB spectrum. The green (light grey) ellipse presents the {pairs} obtained
    using the CMB + SZ spectrum. Right panel: The black solid and green dashed lines are superimposed. They represent the pure CMB
    spectra respectively for the reference cosmological model and a
    one with a shifted value of $\Omega_{\rm dm}h^2$ (see text for
    details). The red thick solid and dashed lines represent the
    corresponding SZ spectra and the blue solid and dot-dashed line
    represent the CMB+SZ spectra.}
\label{fig:Cl_SZbreakdegeneracy}
\end{figure*}

Figure \ref{fig:Cl_SZbreakdegeneracy}b
represents two pure primary CMB spectra calculated for
{the reference cosmological model} (black solid line) and for a model
{in which the value of $\Omega_{\rm dm}h^2$ is the reference value
plus one standard deviation that was obtained using method 3 (red dotted line).}
Since this standard deviation is
much smaller (twice) than those obtained from a pure primary
CMB analysis (see figure \ref{fig:diffmethodexact}), these two CMB
spectra are almost identical, showing the impossibility for a pure CMB
analysis to distinguish between the two cases. The green continuous
and dotted lines (figure \ref{fig:Cl_SZbreakdegeneracy}b) respectively represent the SZ power spectra for the
reference and the shifted models. The change in the $\Omega_{\rm
  dm}h^2$ value significantly affects the amplitude of the SZ
spectrum. As a result, as shown by the blue curves, the CMB+SZ spectra
are different and an analysis based on method 3, that uses the information encoded in the SZ spectrum shape,
allows to disentangle the two cosmological models and consequently better constrains the cosmological parameters
than a pure primary CMB analysis.

We note that the spectral index $n_{\rm s}$ is the only parameter that is somewhat less precisely
constrained through method 3
than it is in methods 1 or 2. An increase of its value {raises both the CMB power} and
SZ spectra at small scales and {lowers their power} at large
scales.  {However,} while the CMB spectrum pivots at intermediate scales,
this happens at large multipoles for the SZ
spectrum. As a result an increase of $n_{\rm s}$ has two
  competing effects between these {domains}~: it increases
  the CMB power and at the same time decreases the SZ power. The
  resulting CMB+SZ spectra for different, but close, values of $n_{\rm s}$
are thus more similar to each other than pure CMB spectra.
  That is why the spectral index is slightly less accurately constrained in
{joint analysis of CMB+SZ than in an analysis of a pure CMB signal.
Nevertheless this slight degradation of the accuracy on $n_{\rm s}$ is negligible
compared to the improvement of the precision of the other parameters.}

\subsection{Effects of the intra-cluster gas physics}
\label{sec:phys} 
As discussed in section \ref{sec:method} , the computation of the TSZ
  power spectrum involves assumptions about the gas physics.
  Either, on the one hand, hydrodynamical-numerical simulations are
  performed with given models for the gas evolution (adiabatic
  cooling, preheating, feedback, etc.), or on the other hand, theoretical
  computations of the power spectrum choose a model for the intra-cluster
  gas distribution as well as scaling relations between the total mass and
  other cluster physical parameters (e.g. temperature). The large diversity
  {of the possible} models describing the cluster gas properties reflects the difficulty to summarise
  in a simple parameterisation the complexity of the intra cluster gas physics.
The cluster physical description {will inevitably} affect the SZ spectrum and consequently
the cosmological parameter estimation.  

In the following, we focus only on the theoretical computations, for which
modifying the cluster physics is straightforward, and we illustrate
the effects {of such modifications} on the cosmological parameters. We restrict our
study
to one single intra-cluster gas distribution model: the $\beta$-profile. We vary the
value of the isothermal gas temperature $T_{\rm e}$ and the index
$\beta$, i.e. the overall amplitude and {steepness} of the density distribution respectively.
{We run an MCMC with method 3 to fit the data created with different values of $\beta$ and
$T_{\rm e}$.}
The resulting one-dimensional parameter distributions are presented in
Fig. \ref{fig:dataphysiquebeta08}.

{An electron distribution with $\beta=0.8$} (i.e. 20\% {
  larger than the reference value}) corresponds to a {more peaked
  individual SZ} profile. This increases the overall SZ power and
shifts the maximum of the SZ spectrum towards smaller scales. Using a
primary CMB plus SZ spectrum calculated using the intra-cluster gas
reference parameterisation to fit such data results in biasing the
estimation of some cosmological parameters, namely $\Omega_{\rm b}h^2$
(0.85 times the expected accuracy (ea)), $\Omega_{\rm dm}h^2$ ($2.6 \times$
ea) and
$n_{\rm s}$ ($0.83 \times$ ea). This corresponds to biases equal
to 2.4 for $\Omega_\Lambda$ and $\Omega_{\rm m}$, 3.3 for $\sigma_8$ and 2.1 for
$H_0$ in terms of the expected precision. A conservative 5\% difference on $\beta$ ($\beta=0.7$), still biases
$\Omega_{\rm DM}h^2$ value at the level of 0.5 times the expected accuracy.
The values of $\Omega_\Lambda$ and $\Omega_m$, $\sigma_8$ and $H_0$ are also
biased at several tenth of the expected precision (0.4, 0.8 and 0.4).

Overestimating the electronic temperature at a 7\% level \citep[{accuracy on the electron
temperature obtained from X-SZ clusters, see}][]{Reese02,Bonamente06} in the data
analysis results also in a biased estimation of some cosmological
parameters, namely $\Omega_{\rm b}h^2$ ($0.6 \times$ ea), $\Omega_{\rm DM}h^2$ ($1.7 \times$
ea)
and $n_{\rm s}$ ($0.5 \times$
ea). This translates to biases of 1.5 for
$\Omega_\Lambda$ and $\Omega_{\rm m}$, 2.2 for $\sigma_8$ and 1.4 for $H_0$.

\begin{figure*}
\includegraphics[width=16cm]{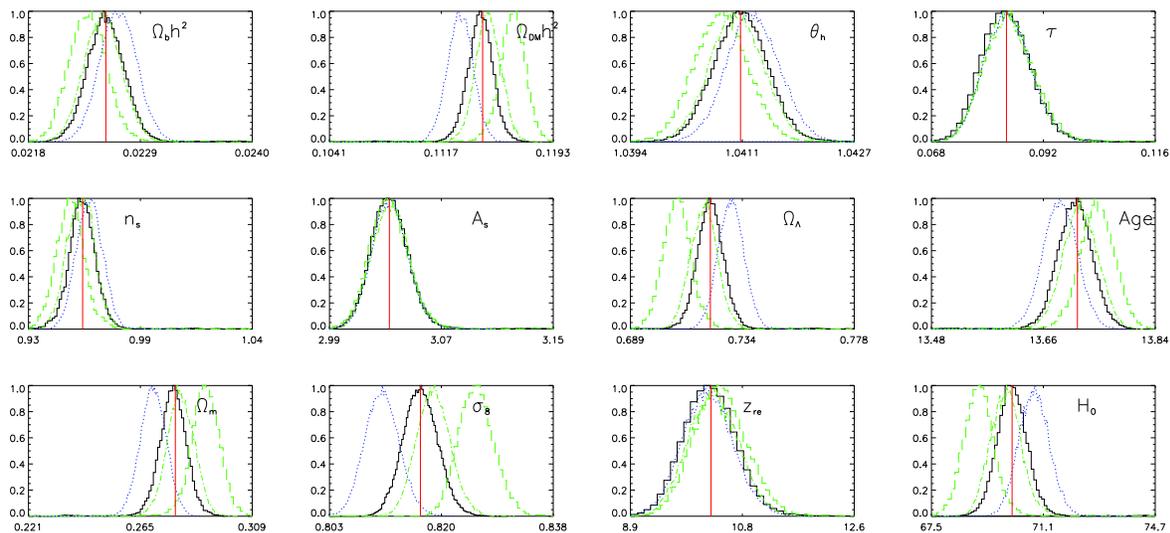}
  \caption{The curves represent the one dimensional parameter
    distribution for the CMB+SZ signal fitted with method 3.
The solid black line represents the parameter distribution
    when the gas description used in the fitting method is in
    agreement with the one employed to create the data. The green
    long-dashed (dot-dashed) line presents the case in which the gas
    description used to create the data uses a $\beta=0.8$ (0.7)
    parameterisation. The blue dotted line stands for the case where the
    electronic temperature is overestimated at a 7\% level in the
    intra cluster gas description used to fit the data. The
    vertical red lines represent the input values of the parameters we
    used to create our {mock} power spectra.}
\label{fig:dataphysiquebeta08}
\end{figure*}

\section{Conclusions}

In this paper we investigate different methods to
jointly fit the primary CMB and SZ signals. We emphasize that
methods using a power spectrum template to describe the SZ
contribution are likely to bias the cosmological parameter values in
the case of an inappropriate choice of the SZ template.
This is due to the frozen shape of the SZ spectrum that
{is inconsistent with changes in} the cosmology.
We also show that a joint CMB-SZ analysis
with a full cosmological dependency of the SZ spectrum does not suffer from
such biases. Moreover it improves the estimate of $\sigma_8$ and
$\Omega_{\rm dm}h^2$ (2.5 and 2 times respectively) with respect to a pure CMB analysis.
In that respect, such a method can be presented as a coherent analysis
of the primary CMB plus SZ signal.
However, we point out that our incomplete understanding of the intra-cluster gas distribution
and properties can result in errors in the calculation of the SZ angular power spectrum
and is likely to bias the cosmological parameter estimation up to 2 sigmas
for $\Omega_{\rm m}$ and $H_0$ and even more for $\sigma_8$.
{This reinforces the need for better constraints on the description
of the galaxy cluster gas properties.}

\section*{Acknowledgments}
The authors thank the referee and they particularly thank
M. Langer for enthusiastic and useful discussions.  We acknowledge the
use of the CAMB and COSMOMC packages.
\label{lastpage}

\bibliographystyle{aa}
\bibliography{biblio}

\begin{thebibliography}{41}
\expandafter\ifx\csname natexlab\endcsname\relax\def\natexlab#1{#1}\fi

\bibitem[{{Bonamente} {et~al.}(2006){Bonamente}, {Joy}, {LaRoque}, {Carlstrom},
  {Reese}, \& {Dawson}}]{Bonamente06}
{Bonamente}, M., {Joy}, M.~K., {LaRoque}, S.~J., {et~al.} 2006, \apj, 647, 25

\bibitem[{{Bond} {et~al.}(2005){Bond}, {Contaldi}, {Pen}, {Pogosyan}, {Prunet},
  {Ruetalo}, {Wadsley}, {Zhang}, {Mason}, {Myers}, {Pearson}, {Readhead},
  {Sievers}, \& {Udomprasert}}]{CBIexcess05}
{Bond}, J.~R., {Contaldi}, C.~R., {Pen}, U.-L., {et~al.} 2005, \apj, 626, 12

\bibitem[{{Bryan} \& {Norman}(1998)}]{BryanNorman98}
{Bryan}, G.~L. \& {Norman}, M.~L. 1998, \apj, 495, 80

\bibitem[{{Cavaliere} \& {Fusco-Femiano}(1976)}]{Cavaliere76}
{Cavaliere}, A. \& {Fusco-Femiano}, R. 1976, \aap, 49, 137

\bibitem[{{Cooray} \& {Melchiorri}(2002)}]{Cooray02alternativsmallscaleexcess}
{Cooray}, A. \& {Melchiorri}, A. 2002, \prd, 66, 083001

\bibitem[{{da Silva} {et~al.}(2001){da Silva}, {Kay}, {Liddle}, {Thomas},
  {Pearce}, \& {Barbosa}}]{daSilva01}
{da Silva}, A.~C., {Kay}, S.~T., {Liddle}, A.~R., {et~al.} 2001, \apjl, 561,
  L15

\bibitem[{{Dawson} {et~al.}(2006){Dawson}, {Holzapfel}, {Carlstrom}, {Joy}, \&
  {LaRoque}}]{BIMAfinal}
{Dawson}, K.~S., {Holzapfel}, W.~L., {Carlstrom}, J.~E., {Joy}, M., \&
  {LaRoque}, S.~J. 2006, \apj, 647, 13

\bibitem[{{Dawson} {et~al.}(2001){Dawson}, {Holzapfel}, {Carlstrom}, {Joy},
  {LaRoque}, \& {Reese}}]{Dawson2001}
{Dawson}, K.~S., {Holzapfel}, W.~L., {Carlstrom}, J.~E., {et~al.} 2001, \apjl,
  553, L1

\bibitem[{{Dolag} {et~al.}(2005){Dolag}, {Hansen}, {Roncarelli}, \&
  {Moscardini}}]{DolagHansen05}
{Dolag}, K., {Hansen}, F.~K., {Roncarelli}, M., \& {Moscardini}, L. 2005,
  \mnras, 363, 29

\bibitem[{{Douspis} {et~al.}(2006){Douspis}, {Aghanim}, \& {Langer}}]{marian06}
{Douspis}, M., {Aghanim}, N., \& {Langer}, M. 2006, \aap, 456, 819

\bibitem[{{Dunkley} {et~al.}(2005){Dunkley}, {Bucher}, {Ferreira}, {Moodley},
  \& {Skordis}}]{Dunkley05}
{Dunkley}, J., {Bucher}, M., {Ferreira}, P.~G., {Moodley}, K., \& {Skordis}, C.
  2005, \mnras, 356, 925

\bibitem[{{Dunkley} {et~al.}(2009){Dunkley}, {Komatsu}, {Nolta}, {Spergel},
  {Larson}, {Hinshaw}, {Page}, {Bennett}, {Gold}, {Jarosik}, {Weiland},
  {Halpern}, {Hill}, {Kogut}, {Limon}, {Meyer}, {Tucker}, {Wollack}, \&
  {Wright}}]{DunkleyWMAP09}
{Dunkley}, J., {Komatsu}, E., {Nolta}, M.~R., {et~al.} 2009, \apjs, 180, 306

\bibitem[{{Elgar{\o}y} {et~al.}(2002){Elgar{\o}y}, {Gramann}, \&
  {Lahav}}]{Elgaroy01}
{Elgar{\o}y}, {\O}., {Gramann}, M., \& {Lahav}, O. 2002, \mnras, 333, 93

\bibitem[{{Griffiths} {et~al.}(2003){Griffiths}, {Kunz}, \&
  {Silk}}]{Griffiths03}
{Griffiths}, L.~M., {Kunz}, M., \& {Silk}, J. 2003, \mnras, 339, 680

\bibitem[{{Holzapfel} {et~al.}(2000){Holzapfel}, {Carlstrom}, {Grego},
  {Holder}, {Joy}, \& {Reese}}]{Holzapfel2000}
{Holzapfel}, W.~L., {Carlstrom}, J.~E., {Grego}, L., {et~al.} 2000, \apj, 539,
  57

\bibitem[{{Jenkins} {et~al.}(2001){Jenkins}, {Frenk}, {White}, {Colberg},
  {Cole}, {Evrard}, {Couchman}, \& {Yoshida}}]{Jenkins01}
{Jenkins}, A., {Frenk}, C.~S., {White}, S.~D.~M., {et~al.} 2001, \mnras, 321,
  372

\bibitem[{{Komatsu} {et~al.}(2009){Komatsu}, {Dunkley}, {Nolta}, {Bennett},
  {Gold}, {Hinshaw}, {Jarosik}, {Larson}, {Limon}, {Page}, {Spergel},
  {Halpern}, {Hill}, {Kogut}, {Meyer}, {Tucker}, {Weiland}, {Wollack}, \&
  {Wright}}]{KomatsuWMAP09}
{Komatsu}, E., {Dunkley}, J., {Nolta}, M.~R., {et~al.} 2009, \apjs, 180, 330

\bibitem[{{Komatsu} \& {Kitayama}(1999)}]{KomatsuKitayama99}
{Komatsu}, E. \& {Kitayama}, T. 1999, \apjl, 526, L1

\bibitem[{{Komatsu} \& {Seljak}(2001)}]{KomatsuSeljak2001}
{Komatsu}, E. \& {Seljak}, U. 2001, \mnras, 327, 1353

\bibitem[{{Komatsu} \& {Seljak}(2002)}]{KomatsuSeljak2002}
{Komatsu}, E. \& {Seljak}, U. 2002, \mnras, 336, 1256

\bibitem[{{Kuo} {et~al.}(2007){Kuo}, {Ade}, {Bock}, {Bond}, {Contaldi}, {Daub},
  {Goldstein}, {Holzapfel}, {Lange}, {Lueker}, {Newcomb}, {Peterson},
  {Reichardt}, {Ruhl}, {Runyan}, \& {Staniszweski}}]{Kuo07}
{Kuo}, C.~L., {Ade}, P.~A.~R., {Bock}, J.~J., {et~al.} 2007, \apj, 664, 687

\bibitem[{{LaRoque} {et~al.}(2006){LaRoque}, {Bonamente}, {Carlstrom}, {Joy},
  {Nagai}, {Reese}, \& {Dawson}}]{Laroque06}
{LaRoque}, S.~J., {Bonamente}, M., {Carlstrom}, J.~E., {et~al.} 2006, \apj,
  652, 917

\bibitem[{{Lewis} \& {Bridle}(2002)}]{LewisBridle02}
{Lewis}, A. \& {Bridle}, S. 2002, \prd, 66, 103511

\bibitem[{{Lewis} {et~al.}(2000){Lewis}, {Challinor}, \& {Lasenby}}]{CAMB2000}
{Lewis}, A., {Challinor}, A., \& {Lasenby}, A. 2000, \apj, 538, 473

\bibitem[{{Liu} {et~al.}(2005){Liu}, {da Silva}, \& {Aghanim}}]{Liu05}
{Liu}, G.-C., {da Silva}, A., \& {Aghanim}, N. 2005, \apj, 621, 15

\bibitem[{{Muchovej} {et~al.}(2007){Muchovej}, {Mroczkowski}, {Carlstrom},
  {Cartwright}, {Greer}, {Hennessy}, {Loh}, {Pryke}, {Reddall}, {Runyan},
  {Sharp}, {Hawkins}, {Lamb}, {Woody}, {Joy}, {Leitch}, \&
  {Miller}}]{Muchovej2007}
{Muchovej}, S., {Mroczkowski}, T., {Carlstrom}, J.~E., {et~al.} 2007, \apj,
  663, 708

\bibitem[{{Padin} {et~al.}(2001){Padin}, {Cartwright}, {Mason}, {Pearson},
  {Readhead}, {Shepherd}, {Sievers}, {Udomprasert}, {Holzapfel}, {Myers},
  {Carlstrom}, {Leitch}, {Joy}, {Bronfman}, \& {May}}]{Padin2001}
{Padin}, S., {Cartwright}, J.~K., {Mason}, B.~S., {et~al.} 2001, \apjl, 549, L1

\bibitem[{{Reese} {et~al.}(2002){Reese}, {Carlstrom}, {Joy}, {Mohr}, {Grego},
  \& {Holzapfel}}]{Reese02}
{Reese}, E.~D., {Carlstrom}, J.~E., {Joy}, M., {et~al.} 2002, \apj, 581, 53

\bibitem[{{Reichardt} {et~al.}(2009){Reichardt}, {Ade}, {Bock}, {Bond},
  {Brevik}, {Contaldi}, {Daub}, {Dempsey}, {Goldstein}, {Holzapfel}, {Kuo},
  {Lange}, {Lueker}, {Newcomb}, {Peterson}, {Ruhl}, {Runyan}, \&
  {Staniszewski}}]{Acbar08}
{Reichardt}, C.~L., {Ade}, P.~A.~R., {Bock}, J.~J., {et~al.} 2009, \apj, 694,
  1200

\bibitem[{{Runyan} {et~al.}(2003){Runyan}, {Ade}, {Bhatia}, {Bock}, {Daub},
  {Goldstein}, {Haynes}, {Holzapfel}, {Kuo}, {Lange}, {Leong}, {Lueker},
  {Newcomb}, {Peterson}, {Reichardt}, {Ruhl}, {Sirbi}, {Torbet}, {Tucker},
  {Turner}, \& {Woolsey}}]{Runyan2003}
{Runyan}, M.~C., {Ade}, P.~A.~R., {Bhatia}, R.~S., {et~al.} 2003, \apjs, 149,
  265

\bibitem[{{Sharp} {et~al.}(2009){Sharp}, {Marrone}, {Carlstrom}, {Culverhouse},
  {Greer}, {Hawkins}, {Hennessy}, {Joy}, {Lamb}, {Leitch}, {Loh}, {Miller},
  {Mroczkowski}, {Muchovej}, {Pryke}, \& {Woody}}]{SZarray09}
{Sharp}, M.~K., {Marrone}, D.~P., {Carlstrom}, J.~E., {et~al.} 2009, ArXiv
  e-prints

\bibitem[{{Sheth} \& {Tormen}(1999)}]{ShethTormen99}
{Sheth}, R.~K. \& {Tormen}, G. 1999, \mnras, 308, 119

\bibitem[{{Sievers} {et~al.}(2009){Sievers}, {Mason}, {Weintraub}, {Achermann},
  {Altamirano}, {Bond}, {Bronfman}, {Bustos}, {Contaldi}, {Dickinson}, {Jones},
  {May}, {Myers}, {Oyarce}, {Padin}, {Pearson}, {Pospieszalski}, {Readhead},
  {Reeves}, {Shepherd}, {Taylor}, \& {Torres}}]{CBI09}
{Sievers}, J.~L., {Mason}, B.~S., {Weintraub}, L., {et~al.} 2009, ArXiv
  e-prints

\bibitem[{{Spergel} {et~al.}(2007){Spergel}, {Bean}, {Dor{\'e}}, {Nolta},
  {Bennett}, {Dunkley}, {Hinshaw}, {Jarosik}, {Komatsu}, {Page}, {Peiris},
  {Verde}, {Halpern}, {Hill}, {Kogut}, {Limon}, {Meyer}, {Odegard}, {Tucker},
  {Weiland}, {Wollack}, \& {Wright}}]{Spergel07}
{Spergel}, D.~N., {Bean}, R., {Dor{\'e}}, O., {et~al.} 2007, \apjs, 170, 377

\bibitem[{{Spergel} {et~al.}(2003){Spergel}, {Verde}, {Peiris}, {Komatsu},
  {Nolta}, {Bennett}, {Halpern}, {Hinshaw}, {Jarosik}, {Kogut}, {Limon},
  {Meyer}, {Page}, {Tucker}, {Weiland}, {Wollack}, \& {Wright}}]{Spergel03}
{Spergel}, D.~N., {Verde}, L., {Peiris}, H.~V., {et~al.} 2003, \apjs, 148, 175

\bibitem[{{Sunyaev} \& {Zel'dovich}(1972)}]{SZ72}
{Sunyaev}, R.~A. \& {Zel'dovich}, Y.~B. 1972, Comments on Astrophysics and
  Space Physics, 4, 173

\bibitem[{{Taburet} {et~al.}(2009){Taburet}, {Aghanim}, {Douspis}, \&
  {Langer}}]{Taburet09}
{Taburet}, N., {Aghanim}, N., {Douspis}, M., \& {Langer}, M. 2009, \mnras, 392,
  1153

\bibitem[{{Warren} {et~al.}(2006){Warren}, {Abazajian}, {Holz}, \&
  {Teodoro}}]{Warren06}
{Warren}, M.~S., {Abazajian}, K., {Holz}, D.~E., \& {Teodoro}, L. 2006, \apj,
  646, 881

\bibitem[{{White} {et~al.}(2002){White}, {Hernquist}, \&
  {Springel}}]{WhiteHernquistSpringel02}
{White}, M., {Hernquist}, L., \& {Springel}, V. 2002, \apj, 579, 16

\bibitem[{{Zhang} \& {Sheth}(2007)}]{ZhangSheth07}
{Zhang}, P. \& {Sheth}, R.~K. 2007, \apj, 671, 14

\bibitem[{{Zhang} \& {Wu}(2003)}]{ZhangWu03}
{Zhang}, Y.-Y. \& {Wu}, X.-P. 2003, \apj, 583, 529

\end{thebibliography}
\end{document}